\documentclass[conference]{IEEEtran}
\IEEEoverridecommandlockouts

\usepackage{cite}
\usepackage{amsmath,amssymb,amsfonts}
\usepackage{graphicx}
\usepackage{textcomp}
\usepackage{xcolor}
\usepackage{url}
\usepackage{booktabs}
\usepackage{tikz}
\usepackage{pgf-umlsd}
\usepackage[disable]{todonotes}
\newif\ifanon
\anonfalse
\newcommand{\anonurl}[1]{\ifanon\emph{[URL omitted for double-blind review]}\else\url{#1}\fi}

\definecolor{colUser}{RGB}{200,220,255}
\definecolor{colPlatform}{RGB}{255,222,173}
\definecolor{colQPU}{RGB}{200,235,200}

\title{Lagrange: Operating Italy's First Publicly-Accessible Quantum Computer for Research and Education\ifanon\else\thanks{This work was supported by the QTech Piemonte strategic initiative.}\fi}

\ifanon
\author{\IEEEauthorblockN{Anonymous Author(s)}
\IEEEauthorblockA{Anonymous Institution(s)\\
Email: anonymous@anonymous.org}}
\else
\author{
\IEEEauthorblockN{Paolo Viviani\IEEEauthorrefmark{1}\IEEEauthorrefmark{4}, Fabrizio Bertone\IEEEauthorrefmark{1}, Giacomo Vitali\IEEEauthorrefmark{1}, Emanuele Dri\IEEEauthorrefmark{1}, Federico Stirano\IEEEauthorrefmark{1},\\
Giuseppe Caragnano\IEEEauthorrefmark{1}, Francesco Lubrano\IEEEauthorrefmark{1}, Marco Cipollini\IEEEauthorrefmark{2}\IEEEauthorrefmark{1}, Martina Giovalli\IEEEauthorrefmark{1},\\
Chiara Vercellino\IEEEauthorrefmark{1}, Alberto Scionti\IEEEauthorrefmark{1}, Virginia Ajani\IEEEauthorrefmark{1}, Antonino Nespola\IEEEauthorrefmark{1}, Olivier Terzo\IEEEauthorrefmark{1},\\
Matteo Cocuzza\IEEEauthorrefmark{2}, Bartolomeo Montrucchio\IEEEauthorrefmark{2}, Giovanna Turvani\IEEEauthorrefmark{2},
Gianluca Bertaina\IEEEauthorrefmark{3}, Marco Coisson\IEEEauthorrefmark{3},\\
Davide Calonico\IEEEauthorrefmark{3}, Fabrizio Pirri\IEEEauthorrefmark{2}, Pietro Asinari\IEEEauthorrefmark{3}}\\
\IEEEauthorblockA{\IEEEauthorrefmark{1}LINKS Foundation, Turin, Italy\\
\IEEEauthorrefmark{2}Politecnico di Torino, Turin, Italy\\
\IEEEauthorrefmark{3}Istituto Nazionale di Ricerca Metrologica (INRiM), Turin, Italy\\
Email:
\IEEEauthorrefmark{4}\{paolo.viviani, fabrizio.bertone, giacomo.vitali, emanuele.dri, federico.stirano,\\
giuseppe.caragnano, francesco.lubrano, martina.giovalli, chiara.vercellino,\\
alberto.scionti, virginia.ajani, antonino.nespola, olivier.terzo\}@linksfoundation.com,\\
\{matteo.cocuzza, marco.cipollini, bartolomeo.montrucchio, giovanna.turvani, fabrizio.pirri\}@polito.it,\\
\{g.bertaina, m.coisson, d.calonico, p.asinari\}@inrim.it}
}
\fi

\begin{document}

\maketitle

\begin{abstract}
We describe the design, implementation, and fourteen-month operational experience of the quantum resource management stack for Lagrange, an IQM Spark five-qubit superconducting quantum computer jointly acquired by LINKS Foundation, Politecnico di Torino and the Italian National Institute of Metrological Research (INRiM), and managed by LINKS. Lagrange is, to our knowledge, the first quantum computer in Italy that is fully operational and accessible to students and researchers from multiple institutions under formal service agreements, and to the general public under commercial agreements. When installed in mid-2025, the IQM Spark hardware was delivered by the vendor with authentication only: no billing, project management or fair usage enforcement were provided. We developed a modular resource management middleware that filled that gap without modifying any vendor client software: a transparent proxy intercepts API calls and enforces project-based QPU-time budgets, reservation-aware allocation, and per-user fairness policies. The middleware adopts a plugin architecture that cleanly separates vendor-specific logic from site-specific policies, enabling reuse across different quantum hardware backends and deployment contexts. Since entering production in September 2025, the system has processed over 850,000 quantum jobs totalling more than 360 hours of QPU execution time. Notably, students at Politecnico di Torino regularly use the machine during both lectures and formal examinations---a practice we believe to be unique in Europe. We report on the system architecture, the operational lessons learned, and the infrastructure choices that made this deployment possible.
\end{abstract}


%%%%%%%%%%%%%%%%%%%%%%%%%%%%%%%%%%%%%%%%%%%%%%%%%%%
\section{Introduction}
\label{sec:introduction}
%%%%%%%%%%%%%%%%%%%%%%%%%%%%%%%%%%%%%%%%%%%%%%%%%%%

The deployment of superconducting quantum computers at research institutions has accelerated in recent years, driven by the commercial availability of machines from vendors such as IQM, IBM, Rigetti, and QuEra. While these machines bring quantum hardware within reach of universities and research centres, their installation entails more than the, nontrivial, facilities-related challenges. In fact, hosting entities typically navigate through a fair amount of different institutional agreements and administrative constraints related to the allocation of resources, or legacy software to be integrated, that lead to software requirements sitting on top of the vendor-provided software. In particular, project-based billing, quota and fairness enforcement, identity federation across partner institutions, and reservation-based scheduling are typically site-specific and require integration with existing infrastructure, thus falling under the responsibility of the hosting site.

Without a functional project management stack, a shared quantum computer is operationally equivalent to a single-user lab instrument: anyone with credentials can submit unlimited jobs, costs cannot be attributed to funded projects, and fair access during high-demand periods cannot be guaranteed. Despite the universality of these needs across quantum computing sites, the published literature offers little guidance on how to address them in practice. Vendor cloud platforms (IBM Quantum, AWS Braket, Azure Quantum) solve these problems internally, but are closed-source, cloud-only, and not deployable on-premises. Academic deployments typically focus on hardware characterisation and benchmarking rather than on the administrative software layer. Moreover, most of these issues involve site-specific details that prevent the portability of existing solutions.

This paper describes our experience building and operating the management software stack for Lagrange, an IQM Spark~\cite{iqm_spark_2024} five-qubit superconducting quantum computer managed by LINKS Foundation, and hosted by Politecnico di Torino in Turin, Italy, within the QCS-Lab (Quantum Computing and Simulation Lab) initiative. The machine was acquired as part of the Regional Strategic Initiative on Quantum Technologies of Piedmont and has been operational since September 2025. It serves researchers, faculty, and students from three partner organisations: LINKS Foundation, Politecnico di Torino, and INRiM (the Italian National Institute of Metrological Research).

To our knowledge, Lagrange is the first quantum computer in Italy that is simultaneously accessible by the general public under commercial agreements, accessible and used by students during regular university courses and formal examinations, and shared across multiple research institutions with heterogeneous billing arrangements. We therefore believe this operational experience to be of value to the broader community of institutions planning similar deployments.

\begin{figure}[h]
  \centering
  \includegraphics[width=\linewidth]{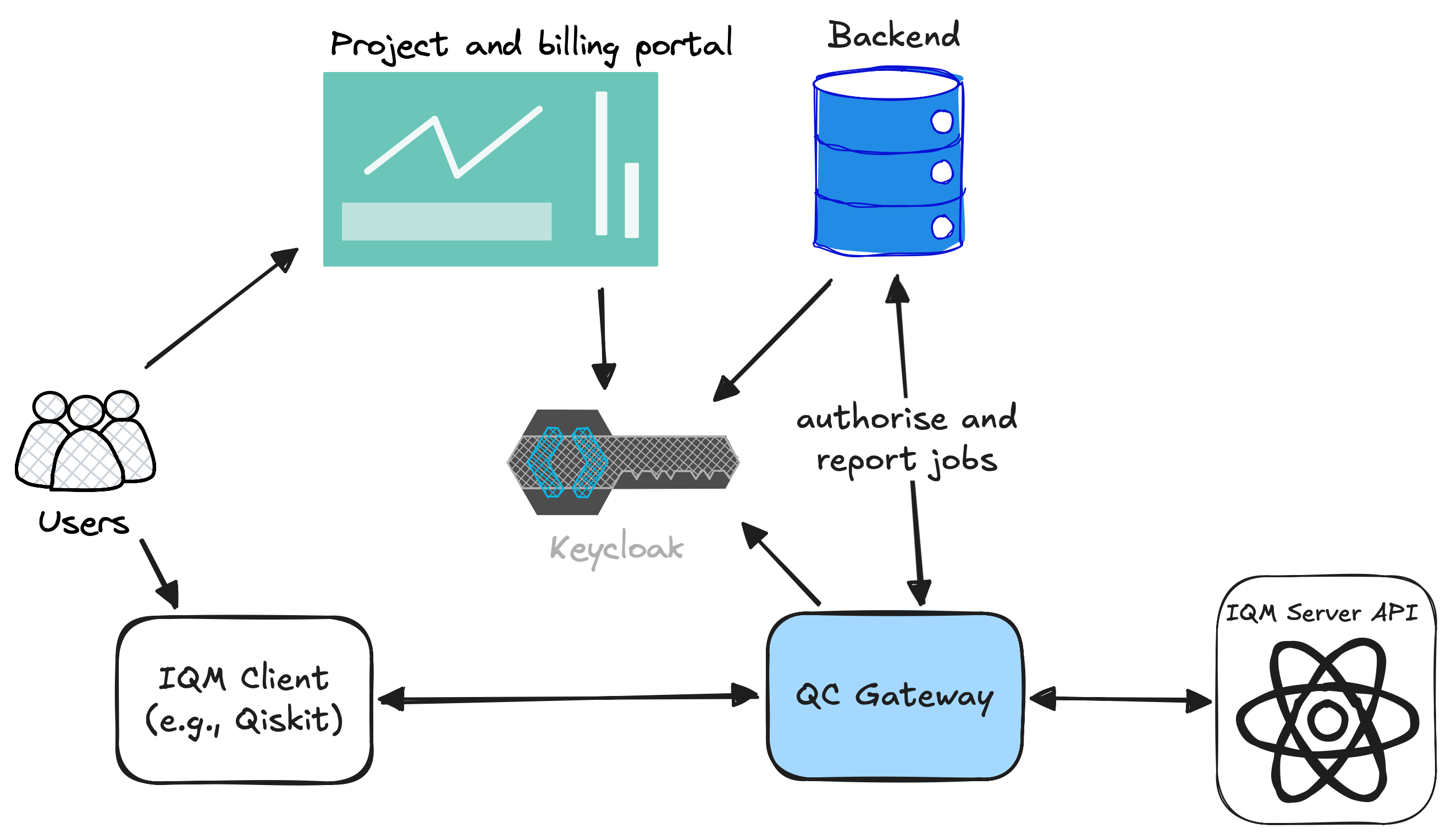}
  \caption{Conceptual architecture of the Lagrange software stack. The QC Gateway transparently authorises job submissions based on budget, reservations, and fair access rate limits. A web portal gives the user visibility to these criteria.}
  \label{fig:concept}
\end{figure}

Our core design principle was \emph{transparency}: the management layer must be invisible to end users. Researchers and students should use unmodified IQM client libraries (Qiskit, Cirq, Qrisp) from their own laptops, exactly as they would against the bare machine or the vendor cloud service. We deliberately rejected a batch-scheduler approach (e.g., SLURM): it is necessary elsewhere to support integration into existing HPC facilities, but would have required SSH access to a head node and fundamentally altered the direct-API workflow that makes quantum computing frameworks productive in both research and education settings. Instead, we built a modular, Python-based middleware that interposes on the IQM REST API as a transparent reverse proxy, adding authorisation, billing, fairness enforcement, and data persistence without changing any request or response scheme. Figure~\ref{fig:concept} outlines the concept behind the software stack we developed. 

A second design principle was \emph{extensibility}. The middleware implements a plugin architecture that separates vendor-specific logic (how to parse IQM API payloads, how to poll job status), and site-specific policies (how to authorise a job against a project budget, how to report consumption to a billing backend), from the core middleware machinery. This separation is baked into the plugin interface, that was designed to support future integration with machines from other vendors and into other hosting sites, including SLURM-based HPC facilities~\cite{vivianihpcqc}.

The remainder of this paper is structured as follows. Section~\ref{sec:background} provides background on the IQM Spark platform and the multi-institutional deployment context. Section~\ref{sec:requirements} formalises the requirements and the assumptions behind the proposed solution. Section~\ref{sec:architecture} describes the system architecture, including the plugin design. Section~\ref{sec:implementation} covers key implementation choices. Section~\ref{sec:infrastructure} describes the hosting infrastructure. Section~\ref{sec:experience} reports on operational experience after more than one year of production use. Section~\ref{sec:related} positions the work relative to related systems. Section~\ref{sec:conclusion} provides closing remarks.

%%%%%%%%%%%%%%%%%%%%%%%%%%%%%%%%%%%%%%%%%%%%%%%%%%%
\section{Background}
\label{sec:background}
%%%%%%%%%%%%%%%%%%%%%%%%%%%%%%%%%%%%%%%%%%%%%%%%%%%

\subsection{IQM Spark Hardware and Software}

The IQM Spark is a five-qubit superconducting quantum computer with star topology. The physical installation comprises a QC Host node (Debian 12), running the Quantum Control Software and the user-facing API server, accessible through a single 10\,Gbps fibre interface via a QC Gateway node, that provides also access to the cryostat control computer and to the remotely-managed power distribution unit.

The machine exposes its functionality through a REST API documented by vendor-provided openapi schemas. In the rest of this paper we will refer to IQM server software (also referred to as IQM OS~\cite{iqm_docs_2026}) v4.4.6, which is the version installed at the time of writing.

The primary endpoint for job submission is \texttt{POST /jobs/\{type\}/circuit}. The official IQM SDK~\cite{iqmsdk} wraps these endpoints in a Python client (\texttt{iqm-client}) that is itself integrated into IQM-compatible versions of Qiskit~\cite{qiskit}, Cirq~\cite{cirq}, and Qrisp~\cite{qrisp}. Job submission is asynchronous: the client posts a circuit definition in terms of native gate operations and receives a \texttt{job\_id}, which is then polled until results (per-shot measurement outcomes for all requested qubits) are available. Concurrent submissions are handled by the machine's internal first-come-first-served queue.

At installation time, authentication was provided via a Keycloak instance on the QC Host, with the \texttt{cortex-cli} tool\footnote{\url{https://github.com/iqm-finland/cortex-cli}} that handled the authentication token flow, writing a \texttt{tokens.json} file read by \texttt{iqm-client}. Critically, \emph{any user with a valid Keycloak account can submit an arbitrary number of jobs with no further restriction} without any billing, quota, or project-based enforcement. IQM OS 4.4.x deprecated keycloak in favor of another custom solution that provides the token through a webpage, and also provides some degree of reservation enforcement capabilities, but no project-level budget. The machine also exposes calibration data endpoints, providing a rich set of information related to the various performance metrics of the quantum computer.

\subsection{Multi-Institutional Context}
\label{sec:context}

The deployment involves three partner organisations with distinct access and billing agreements:

\textbf{LINKS Foundation} is the machine owner and administrator, responsible for day-to-day operation, maintenance scheduling, and access regulation. LINKS users are typically internal researchers or R\&D personnel from partner institutions.

\textbf{Politecnico di Torino} physically hosts the machine and co-owns usage rights with LINKS. Its user base is the most diverse: it spans from faculties to large student cohorts who use the machine as part of quantum computing courses. The student use case is particularly interesting: it involves dozens of simultaneous users submitting jobs within narrow time windows during laboratory sessions and, notably, during formal examinations.

\textbf{INRIM} (Italian National Institute of Metrological Research) holds a time-purchase agreement and receives a pre-defined allocation of machine time, which is partially shared with external collaborators.

Additionally, external users from other institutions may receive time allocations through contracts with either LINKS or Politecnico, for example research collaborators on funded European projects. The heterogeneity of these agreements---combining pre-allocated slot ownership, project-based budget access, exclusive reservations, and best-effort open queue access---imposes a rich and challenging set of requirements to the middleware development, being the primary driver of the system's complexity.

%%%%%%%%%%%%%%%%%%%%%%%%%%%%%%%%%%%%%%%%%%%%%%%%%%%
\section{Requirements}
\label{sec:requirements}
%%%%%%%%%%%%%%%%%%%%%%%%%%%%%%%%%%%%%%%%%%%%%%%%%%%

Through a structured requirements-gathering process involving all partner institutions, conducted before implementation began, we identified the following categories of requirements.

\paragraph{Access}
The system must remain fully compatible with unmodified \texttt{iqm-client}, Qiskit, Cirq, and Qrisp: users should not need to change their workflow (\emph{client transparency}). Users must be able to submit jobs directly from a laptop without SSH access to a head node, explicitly excluding SLURM as the primary access mechanism (\emph{no batch scheduler}).

\paragraph{Access and Identity}
Users must authenticate with a single set of credentials across all system components (\emph{SSO}). LINKS and Politecnico users must be able to log in with their institutional credentials via federated identities. 

\paragraph{Accounting and Billing}
Every job must be associated with a project, either explicitly via metadata in the submission or implicitly via a user's configured default project (\emph{project-based billing}). Each project has a budget measured in QPU milliseconds. Job submission must be rejected if the project has exhausted its budget (\emph{budget enforcement}). The system must generate accounting reports for each organisation and project (\emph{billing reports}). A per-user limit on simultaneously active jobs must be enforced (\emph{fairness}).

\paragraph{Resource Allocation}
LINKS and Politecnico can receive blocks of machine time in advance (\emph{pre-allocated time slots}), reflecting their bilateral agreement to split guaranteed availability time equally. Within a pre-allocated slot, the owning organisation may create \emph{exclusive reservations}, preventing other users from submitting jobs. Outside reserved periods, the machine must be accessible through a fair FIFO queue (\emph{open queue access}). Unused time in a pre-allocated slot can be used by the other party with queue access. Time outside the guaranteed availability window is accessible on a \emph{best-effort} basis.

\paragraph{Long-term data storage and control}
Submitted circuits and job results must be stored persistently and made accessible to the submitting user, independently of IQM's internal storage (\emph{long-term result storage}). The calibration report in effect at the time of each job submission must be archived and associated with the job record (\emph{calibration archiving}), enabling users to inspect the machine performance metrics related to their job submission. Long-term storage must be also ensured for the machine's operational metrics collected through Prometheus.

\paragraph{Administrative Interface}
Administrators and users require a web dashboard to manage organisations, projects, users, budgets, reservations, and to browse job history and results. Real-time and historical machine status monitoring must be also reported.

%%%%%%%%%%%%%%%%%%%%%%%%%%%%%%%%%%%%%%%%%%%%%%%%%%%
\section{System Architecture}
\label{sec:architecture}
%%%%%%%%%%%%%%%%%%%%%%%%%%%%%%%%%%%%%%%%%%%%%%%%%%%

% \begin{figure*}[h]
%   \centering
%   \includegraphics[width=0.85\textwidth]{sw_architecture.png}
%   \caption{Software architecture of the Lagrange management stack. User requests enter through three client interfaces (Python/Qiskit code, \texttt{lagrangeclient} CLI, or browser) and are routed by an Nginx reverse proxy to the appropriate backend services. The API middleware transparently intercepts IQM API calls, enforcing authorisation and billing before forwarding to the quantum computer. Storage services (PostgreSQL, Redis, RustFS) provide persistence, concurrency control, and long-term result archiving.}
%   \label{fig:arch}
% \end{figure*}

% Figure~\ref{fig:arch} shows the full software architecture. The system is organised into four layers: client interface, routing, service, and storage.

\subsection{Client Interface Layer}
Users interact with the system through three entry points: Python code, a CLI tool, and a browser interface. The primary mode is Python code (e.g., Qiskit, Cirq or Qrisp), although C++ with CUDA-Q is also supported by the vendor: a user writes a quantum program, sets the \texttt{IQM\_SERVER\_URL} environment variable to point to our infrastructure frontend, and submits jobs programmatically with no code modifications required. A lightweight CLI tool, \textbf{\texttt{lagrangeclient}}, replaces the vendor-provided \texttt{cortex-cli}, pointing to a LINKS-managed Keycloak instance to handle the SSO authentication flow and write a \texttt{tokens.json} file compatible with \texttt{iqm-client}. Finally, a \textbf{browser} interface provides access to the projects dashboard, billing portal, reservation calendar, job results, and Perses monitoring dashboards.

\subsection{Routing Layer}
An Nginx reverse proxy with TLS termination (via Let's Encrypt) sits in front of all services, implementing subdomain-based routing as shown in Table~\ref{tab:nginx_routing}. This architecture keeps all components behind a single entry point, simplifying certificate management and firewall rules.

\begin{table}[h]
\centering
\caption{Nginx reverse proxy subdomain routing configuration}
\label{tab:nginx_routing}
\begin{tabular}{ll}
\toprule
\textbf{Subdomain} & \textbf{Service} \\
\midrule
\texttt{spark.*} & QC API middleware \\
\texttt{spark.*/auth} & Keycloak authentication \\
\texttt{api.*} & Backend API \\
\texttt{dashboard.*} & Project management portal \\
\texttt{grafana.*} & Perses monitoring tool \\
\texttt{store.*} & S3-compatible object store \\
\texttt{jobs.*} & Job results visualization interface \\
\bottomrule
\end{tabular}
\end{table}

\subsection{Service Layer}
\paragraph{QC Gateway}
QC Gateway\footnote{\anonurl{https://github.com/LINKS-Foundation-CPE/QC-Gateway}} is an HTTP middleware implemented in Python using FastAPI. 
It acts as a transparent proxy for all IQM API calls. When a client submits a job via \texttt{POST /jobs/\{type\}/circuit}, the middleware performs the following steps:

\begin{enumerate}
  \item Validates the OIDC bearer token against the LINKS Keycloak instance.
  \item Performs role-based access control: verifies that the user's role permits the requested API operation; access to the pulse-level endpoint is further restricted to users for whom an administrator has enabled a dedicated flag in the dashboard.
  \item Calls the \texttt{/jobAuthoriser} endpoint on the backend API, passing user identity and requested project.
  \item Checks the per-user concurrent shot limit via Redis (fairness quota enforcement).
  \item If all checks pass, forwards the request to the IQM Server API with a service token, records the returned \texttt{job\_id}, and returns the response to the client unchanged.
  \item Reports the successful job submission to the backend API at the \texttt{/jobReporter} endpoint.
  \item On job completion, the background job reporter updates the concurrency quota, then calls the \texttt{/jobReporter} endpoint to update budget consumption and uploads the circuit definition, results, and calibration metrics to the object store.
\end{enumerate}

\begin{figure*}[!h]
\centering
\scalebox{0.75}{%
\small
\begin{sequencediagram}
  % --- Local Frontend ---
  \tikzstyle{inststyle}=[rectangle, draw, anchor=west,
    minimum height=0.8cm, minimum width=1.6cm,
    fill=colUser, drop shadow={opacity=0.5}]
  \newthread[colUser]{user}{User (IQM client)}
  \newinst{cli}{\texttt{lagrangeclient}}

  % --- Platform Backend ---
  \tikzstyle{inststyle}=[rectangle, draw, anchor=west,
    minimum height=0.8cm, minimum width=1.6cm,
    fill=colPlatform, drop shadow={opacity=0.5}]
  \newinst[1.]{kc}{Keycloak}
  \newinst{mw}{QC Gateway}
  \newinst{be}{Backend API}
  \newinst{rd}{Redis}
  \newinst{s3}{Object Store}
  \newthread[colPlatform]{jr}{Job Reporter}

  % --- External QPU ---
  \tikzstyle{inststyle}=[rectangle, draw, anchor=west,
    minimum height=0.8cm, minimum width=1.6cm,
    fill=colQPU, drop shadow={opacity=0.5}]
  \newinst[1.]{qc}{IQM Spark}
  
  % === All sdblocks ===
  \begin{sdblock}{Authentication (one-time)}{}
    \begin{call}{user}{\texttt{login}}{cli}{\texttt{tokens.json}}
      \begin{call}{cli}{OIDC flow}{kc}{access token}
      \end{call}
    \end{call}
  \end{sdblock}
  \begin{sdblock}{Job Submission and status polling}{}
    \begin{call}{user}{\texttt{POST /jobs/circuit}}{mw}{\texttt{job\_id}}
      \begin{call}{mw}{validate token}{kc}{identity,roles}
      \end{call}
      \begin{call}{mw}{\hspace{110pt}\texttt{/jobAuthoriser(user, project)}}{be}{authorised}
      \end{call}
      \begin{call}{mw}{check concurrency}{rd}{OK (incremented)}
      \end{call}
      \begin{call}{mw}{forward request (using service account token)\hspace{30pt}\hphantom{a}}{qc}{\texttt{job\_id}}
      \end{call}
      \begin{call}{mw}{upload circuit (async)}{s3}{}
      \end{call}
      \begin{call}{mw}{\hspace{60pt}\texttt{/jobReporter} (submitted)}{be}{}
      \end{call}
    \end{call}
    \begin{call}{user}{\texttt{GET /jobs/\{job\_id\}}}{mw}{status}
      \begin{call}{mw}{forward query}{qc}{status}
      \end{call}
    \end{call}
  \end{sdblock}
  \begin{sdblock}{Background Job Monitoring}{}
    \begin{call}{jr}{poll status}{qc}{ready + results}
    \end{call}
    \begin{call}{jr}{upload results and calibration\hspace{70pt}\hphantom{a}}{s3}{}
    \end{call}
    \begin{call}{jr}{\texttt{/jobReporter} (completed, QPU time)}{be}{}
    \end{call}
    \begin{call}{jr}{decrement concurrency counter\hspace{30pt}\hphantom{a}}{rd}{}
    \end{call}
  \end{sdblock}
\end{sequencediagram}%
}% end scalebox

\vspace{-0.5em}
% Legend below the diagram, inside the figure
\centering
\footnotesize
\begin{tikzpicture}
  \node[draw=gray!50, rounded corners=3pt, inner sep=6pt, fill=white] {
    \tikz\draw[fill=colUser, draw=gray] (0,0) rectangle (0.35,0.25);
    \,User Machine\qquad
    \tikz\draw[fill=colPlatform, draw=gray] (0,0) rectangle (0.35,0.25);
    \,Access Platform\qquad
    \tikz\draw[fill=colQPU, draw=gray] (0,0) rectangle (0.35,0.25);
    \,Lagrange QPU
  };
\end{tikzpicture}

\caption{Sequence diagram of the job submission and completion flow.}
\label{fig:sequence}
\end{figure*}
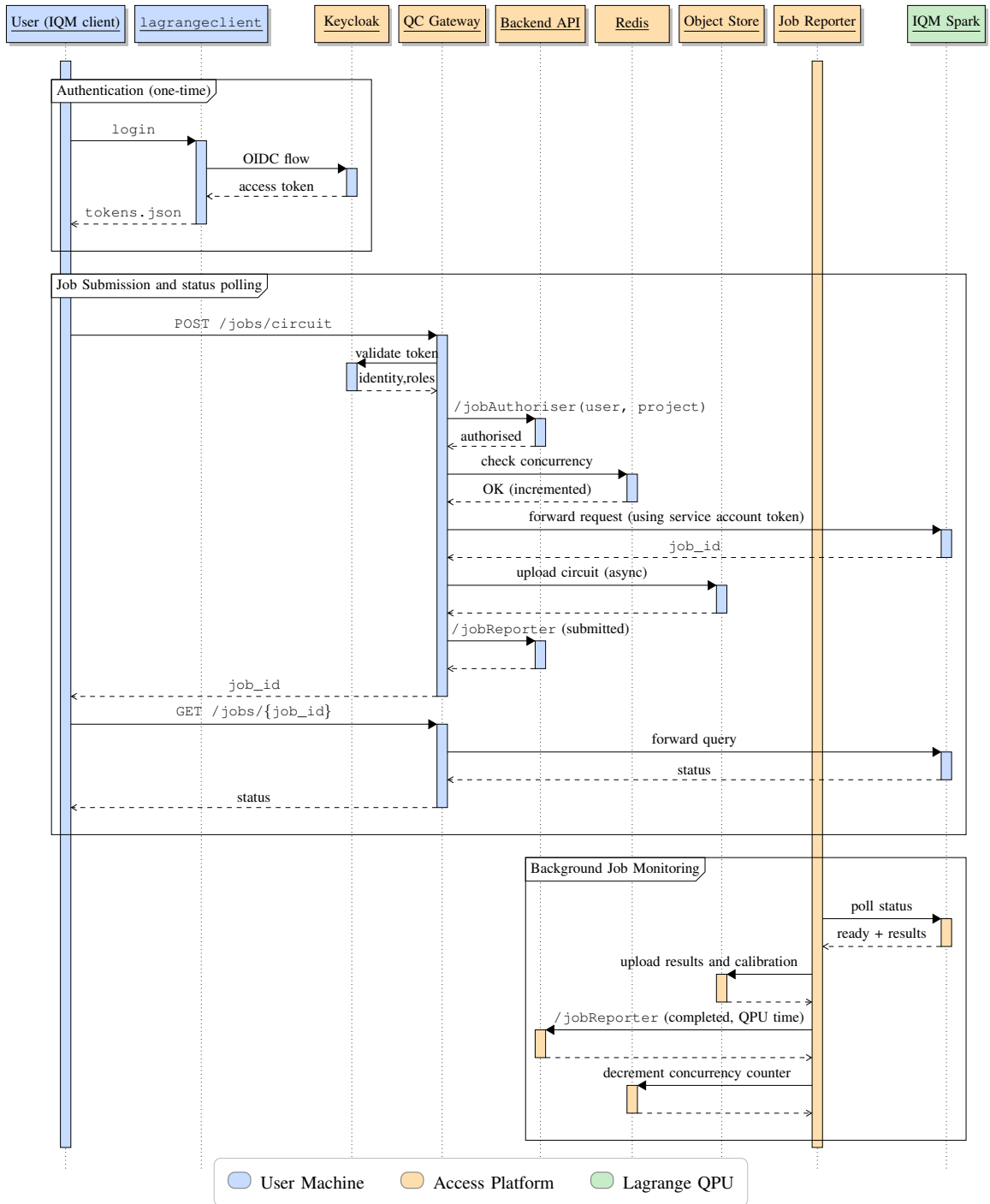

\noindent Figure~\ref{fig:sequence} illustrates the complete flow, including authentication, submission, polling, and background completion processing.

The proxy preserves all API paths, request/response schemas, and HTTP status codes. Non-submission endpoints (health checks, calibration queries) are forwarded without modification. This full transparency means that \texttt{iqm-client} and any framework integrating it, including Qiskit and Cirq, work without any modification.

The middleware is designed to be \emph{fail-open} for non-critical services: if Redis or RustFS are temporarily unavailable, the proxy continues to forward jobs to the machine. Concurrency checks and result uploads are retried asynchronously. This ensures that transient infrastructure issues do not block researchers from using the quantum computer.

\paragraph{Plugin Architecture}
\label{sec:plugins}

A key architectural decision is the separation of the middleware's generic proxy logic from vendor-specific and site-specific concerns through a plugin system. The middleware defines two plugin interfaces:

\textbf{Vendor plugins} encapsulate everything specific to a quantum computer vendor's API. It was specifically designed so that the same middleware core can be reused with different quantum hardware backends by implementing only the vendor-specific request parsing and polling logic: route definitions (which API paths exist and which roles they require), request parsing (extracting shot counts, circuit counts, and project metadata from submission payloads), response parsing (extracting job IDs), header construction (injecting vendor-specific authentication into upstream requests), job status polling, job artifact classification, and calibration handling.

\textbf{Site plugins} encapsulate how a specific hosting site handles the access to the machine, enabling QC Gateway to operate under different hosting site policies including job authorisation (checking whether a user is allowed to submit given site-specific budget rules), and job reporting (sending status updates to the billing backend), and result URL construction. The site plugin interface would allow, for instance, to replace the Lagrange billing backend with a SLURM accounting system for HPC-centre deployments.

Both plugin types communicate with the core through well-defined dataclasses (\texttt{JobSubmission}, \texttt{SubmissionResult}, \texttt{JobStatusResult}, \texttt{JobAuthorizationResult}, etc.), ensuring a clean separation of concerns.

The current deployment uses an IQM vendor plugin and a \emph{Lagrange} site plugin. Plugin selection is configured via environment variables (\texttt{VENDOR\_PLUGIN}, \texttt{SITE\_PLUGIN}), and plugins are loaded dynamically at application startup through a registry.

\paragraph{Backend API}
The backend API is specific for the Lagrange site deployment, and manages the persistent state of the system. Its data model includes the resources listed in Table~\ref{tab:resource_hierarchy}.
\begin{table}[h]
\centering
\caption{Resource hierarchy and access control model}
\label{tab:resource_hierarchy}
\begin{tabular}{lp{6cm}}
\toprule
\textbf{Entity} & \textbf{Description} \\
\midrule
Organisations & Top-level entities corresponding to partner institutions (LINKS, Politecnico, INRiM). Each is assigned a yearly QPU-time budget and pre-allocated time slots. \\
\midrule
Projects & Funding entities within an organisation with their own QPU-time budget drawn from the parent allocation. Support cross-institutional collaborations with multiple users from different organisations. Managed by Project Administrators (PIs) who can add collaborators and create exclusive reservations. \\
\midrule
Users & Individuals belonging to one or more organisations and projects. Have role-based permissions (admin, organisation manager, PI, regular user), a default project for job billing, and can define their own default project. \\
\midrule
Pre-allocated slots & Time windows during which each organization can place reservations. \\
\midrule
Reservations & Time windows during which the machine is exclusively accessible to a specific project's users. \\
\midrule
Jobs & Records linking job IDs to users, projects, timestamps, shot counts, and QPU time consumed, with pointers to long-term result storage. \\
\bottomrule
\end{tabular}
\end{table}

% \textbf{Organisations and sub-organisations}: top-level entities corresponding to partner institutions (LINKS, Politecnico, INRiM). Each organisation is assigned a yearly QPU-time budget and a set of pre-allocated time slots.

% \textbf{Projects}: funding entities within an organisation. Projects carry a QPU-time budget drawn from the parent organisation's allocation. Multiple users from multiple organisations can belong to a project, supporting cross-institutional collaborations. Projects have one or more Project Administrators (PIs) who can add collaborators and create exclusive reservations.

% \textbf{Users}: individuals who belong to one or more organisations and projects, with role-based permissions (admin, organisation manager, PI, regular user). Each user has a default project to which jobs are billed when no explicit project is specified, and that can be defined by the users themselves.

% \textbf{Reservations}: time windows during which the machine is exclusively accessible to a specific project's users.

% \textbf{Jobs}: records linking job IDs to users, projects, timestamps, shot counts, and QPU time consumed. It also carries a pointer to the long-term storage of the results.

The backend exposes two endpoints consumed by the QC Gateway (\texttt{/jobAuthoriser} and \texttt{/jobReporter}) and a richer REST API consumed by the dashboard.

\paragraph{Projects Dashboard}

The dashboard is a web application that provides a management interface for projects, combined with a calendar application, and a job monitor. Key views include a \emph{slots calendar} showing pre-allocated time blocks per organisation; a \emph{reservations calendar} for viewing and creating exclusive reservations within such time blocks; a \emph{jobs table} displaying submitted circuits, measurement results, and histograms; and a \emph{project panel} summarising used and remaining QPU time per project. Users can also access the calibration report associated with each of their jobs directly from the dashboard.

\paragraph{Authentication and Federation}

We operate our own Keycloak instance as the central identity provider, towards which the middleware authenticate requests. Authentication towards the QC is managed with a dedicated service user.

Federation with Politecnico di Torino's identity infrastructure is implemented via SAML-based identity brokering in Keycloak. LINKS internal users authenticate via the institutional GitLab as an identity provider, federated with Keycloak via OpenID Connect (OIDC). For Politecnico's student users, federation enables seamless login with institutional credentials, overcoming the need to manually provision hundreds of accounts. Users in the INRiM organization are also authenticated via the LINKS GitLab.

\subsection{Storage Layer}

The storage layer comprises four components.
\begin{itemize}
  \item A \textbf{PostgreSQL database} stores all persistent state for the backend API: organisations, projects, users, roles, pre-allocated slots, reservations, and billing records.
  \item A separate, lightweight \textbf{PostgreSQL instance} local to the QC Gateway tracks currently active jobs, enabling status tracking and reporting; jobs are deleted from this DB once they reach a terminal status.
  \item A \textbf{Redis instance} enforces per-user concurrent job limits via atomic counters: before forwarding a submission, the middleware increments a counter; the counter is decremented upon job completion; submissions exceeding the limit are rejected with HTTP~429 (Too Many Requests).
  \item An \textbf{S3-compatible object store} (RustFS) provides safe, long-term storage for submitted circuit definitions, job results, and calibration reports, keyed by job ID. This represents a significant safeguard from the machine operator's perspective, as it does not rely on the machine's own internal storage, which the operator has little visibility on.
\end{itemize}

%%%%%%%%%%%%%%%%%%%%%%%%%%%%%%%%%%%%%%%%%%%%%%%%%%%
\section{Implementation Details}
\label{sec:implementation}
%%%%%%%%%%%%%%%%%%%%%%%%%%%%%%%%%%%%%%%%%%%%%%%%%%%

\subsection{Authorisation Logic}

The authorisation check, implemented in the site plugin's \texttt{authorize\_job} method and the backend's \texttt{/jobAuthoriser} endpoint, encodes the full complexity of the multi-layered access model:

\begin{enumerate}
  \item Is the user authenticated and a valid member of any project? If not, reject.
  \item Is there a current active exclusive reservation? If yes, does the reservation belong to the user's project? If not, reject.
  \item Does the user's selected project have sufficient remaining budget? If not, reject. (Jobs submitted during a projects reservation do not consume budget, as it was already consumed when placing the reservation.)
  \item Does the submission respect the fair use limits for the given user (c.f. Section~\ref{sec:fairness})? If not, reject.
  \item Approve.
\end{enumerate}

\subsection{Fairness Enforcement}
\label{sec:fairness}

Beyond project-level budget enforcement, the system imposes hard per-user concurrency limits to prevent any single user from flooding the machine queue. The limit is expressed in terms of total outstanding shots across all concurrent jobs, plus the one being submitted:

\begin{equation}
\sum_{j=1}^{J} \texttt{num\_circuits}_j \cdot \texttt{shots}_j \leq S_{\max}
\label{eq:fairness}
\end{equation}

\noindent where $J$ is the number of jobs currently submitted (including the one being submitted) by the user that have not yet reached a terminal state (\texttt{ready} or \texttt{failed}), and $S_{\max}$ is a configurable threshold (currently set to 2.5 million shots, corresponding to approximately five minutes of QPU time). When the limit is reached, the user must wait for an active job to complete before submitting a new one. In practice, the shot count has proven to be only a rough proxy for QPU time: with longer circuits, the outstanding workload admitted by this threshold can significantly exceed five minutes, so $S_{\max}$ is likely to be retuned, and work is ongoing to identify a better proxy for QPU time.

The concurrency state is maintained in Redis via atomic increment and decrement operations, keyed by username. The middleware implements rollback logic: if the upstream submission fails after the counter has been incremented, the counter is rolled back to prevent phantom reservations.

\subsection{Login Client}

We implemented \texttt{lagrangeclient}~\cite{lagrangeclient}, a thin Python CLI that replaces \texttt{cortex-cli}. It performs the OIDC device flow or browser-redirect flow against our Keycloak instance and writes a \texttt{tokens.json} file in the format expected by \texttt{iqm-client}. This is the only component that end users need to install beyond the standard IQM SDK. The tool is distributed as a Python package and can be installed via \texttt{pip}.

\subsection{Background Job Reporter}

A background daemon (\texttt{job\_reporter.py}) continuously polls the IQM machine
for the status of active jobs as they are reported in the local PostgreSQL database. For each job that reaches a terminal state, the reporter:
\begin{enumerate}
    \item Fetches the job artifacts (results, timeline) via the vendor plugin;
    \item Uploads artifacts and the calibration report current at submission time to RustFS;
    \item Builds a user-facing result URL;
    \item Sends a completion report to the backend via the site plugin's \texttt{report\_job\_sync} method;
    \item Decrements the Redis concurrency counter.
\end{enumerate}
The reporter also handles vendor-specific calibration report fetching and storage. The archived calibration reports also form a user-accessible historical record of the machine's calibration data.

\subsection{Containerised Deployment}

All services are containerised with Docker and orchestrated via Docker Compose. Nginx, the API middleware, the backend API, Keycloak, the dashboard, Perses, Prometheus, Loki, Redis, RustFS, and the databases each run as independent containers. This approach simplifies upgrades, migrations and keeps the overall DevOps effort compatible with the workload of the research staff involved. 

%%%%%%%%%%%%%%%%%%%%%%%%%%%%%%%%%%%%%%%%%%%%%%%%%%%
\section{Hosting Infrastructure}
\label{sec:infrastructure}
%%%%%%%%%%%%%%%%%%%%%%%%%%%%%%%%%%%%%%%%%%%%%%%%%%%

All frontend software components---the API middleware, backend API, dashboard, Keycloak, monitoring stack, and object store---are hosted on LINKS Foundation's development cloud infrastructure. This is a private OpenStack cluster providing approximately 400 CPU cores, 1.5\,TB of RAM, 6 GPUs, and 90\,TB of NVMe-backed Ceph storage. The quantum computer's Gateway node is connected to this infrastructure via the campus' 10\,Gbps fibre link.

The entire software stack is managed through CI/CD pipelines (GitLab CI) that automate building, testing, and deployment both on the production environment and on a development environment dedicated to integration testing. This environment is composed of a Virtual Machine (VM) hosting an IQM mock device that replicates the full IQM API but returns meaningless results, and a development node running the whole frontend software stack that regulates access to the mock device. This development environment is also used to validate that the frontend software remains compatible with software updates from the vendor.

Observability is provided by a dual-purpose monitoring stack based on Perses, Prometheus, and Loki. The first function is \emph{QC telemetry mirroring}: the monitoring stack mirrors data from the quantum computer's internal Prometheus instance, exposing dashboards for public access. This allows users and administrators to inspect real-time and historical machine health data, including qubit fidelity metrics and cryostat diagnostics, without direct access to the machine's internal network. The second function is \emph{software stack observability}: Prometheus collects metrics and Loki aggregates structured logs from all software stack components, enabling rapid diagnosis of issues.

%%%%%%%%%%%%%%%%%%%%%%%%%%%%%%%%%%%%%%%%%%%%%%%%%%%
\section{Operational Experience}
\label{sec:experience}
%%%%%%%%%%%%%%%%%%%%%%%%%%%%%%%%%%%%%%%%%%%%%%%%%%%

The system has been in continuous production since September 2025. As of November 2026, it has processed over 850,000 quantum jobs totalling more than 360 hours of actual QPU execution time. In this timeframe it served about 1,000 users from the participating institutions and their partners, including about 750 students. Lagrange was also used to run what is, to our knowledge, the first hybrid HPC-QC job fully executed in Italy, coupling the Leonardo supercomputer with Lagrange; this experiment, together with the HPC-QC scheduling strategies it was used to evaluate, is described in~\cite{CIPOLLINI2026108699}.

\subsection{Educational Activities}
\label{sec:education}

A distinctive aspect of the Lagrange deployment is its integration into advanced teaching activities, where access to a real quantum processor is embedded within a highly interactive educational model. Students operate in small groups and are required to apply theoretical concepts through hands-on experimentation, not only via simulation tools but also through direct interaction with the quantum hardware.

These activities are developed within a recently established Master's degree programme in Quantum Engineering\footnote{\url{https://www.polito.it/en/education/master-s-degree-programmes/quantum-engineering}}, which aims at training engineers capable of operating across the full spectrum of quantum technologies, including computing, communication, and sensing. The programme is inherently multidisciplinary, combining elements of physics, electronic engineering, and computer science, and is designed to prepare students for both research-oriented and industrial applications of quantum technologies.

This educational setting naturally involves cohorts with heterogeneous backgrounds, including physics, electronic engineering, computer engineering, and information engineering. The diversity of competencies enables a collaborative learning process in which different perspectives converge on the design, implementation, and validation of quantum circuits and systems.

Students typically work in small teams, where they are tasked with translating theoretical constructs into executable experiments, analysing discrepancies between ideal and real-device behaviour, and interpreting results in the presence of noise and hardware constraints. The availability of an on-premises quantum computer allows these activities to be tightly integrated with lectures and laboratory sessions, enabling iterative experimentation and immediate validation of design choices.

The teaching activities developed around Lagrange can be broadly classified into two complementary perspectives: a software- and application-oriented usage, where the system is employed as an execution backend for algorithms and programming exercises, and a hardware- and system-oriented usage, where the quantum processor is treated as a physical platform for engineering analysis, control, and optimisation.

\subsubsection{Software-Oriented Educational Activities}
\label{sec:education_software}

Starting from October 2025, teaching activities involving Lagrange have been introduced in several courses.

In a \textit{Quantum Computing} course in the second year of a Master's degree programme in Quantum Engineering, the Lagrange computer has been used to demonstrate algorithms and support exercises during lectures and laboratory sessions. All students have used the system for their individual exercises, accessing with their institutional accounts and with a project dedicated to the course, with generous budget accessible at any time.

Lagrange has also been used during the final examination of the course, where students executed the examples required by the exam assignments on the quantum processor. The institutional e-learning platform is used to manage submissions and results.

The system has also been used in other teaching contexts, including training activities involving industrial partners.
Finally, Lagrange is extensively used by QubiTO\footnote{\url{https://qubito.polito.it/en/}}, the first quantum computing student team in Italy, whose members participated in hackathons around Europe and organized events focused on quantum technologies.

\subsubsection{Hardware-and System-Oriented Educational Activities}
\label{sec:education_hardware}

In parallel with software-oriented usage, Lagrange enables another class of teaching activities focused on the hardware and system-level aspects of quantum computing. These activities are currently being structured within advanced courses on \emph{Qubit Electronics} and \emph{Quantum Hardware Design and Optimization}, forming a coherent framework that connects device-level physics with system-level design and optimisation.

The \emph{Qubit Electronics} course addresses the physical and electronic implementation of qubits, including coherence mechanisms, noise sources, and control and readout infrastructures. In contrast, the \emph{Quantum Hardware Design and Optimization} course focuses on circuit design, compilation strategies, hardware-aware optimisation, and the integration of quantum algorithms with classical computing platforms.

Within this context, a first class of activities targets the physical understanding of qubits and their electronic environment. Students analyse superconducting qubits as engineered devices, studying coherence properties, decoherence mechanisms, and the impact of control electronics and environmental coupling. These aspects are complemented by interaction with the quantum processor, enabling the comparison between theoretical models and observed device behaviour.

A second class of activities focuses on the full control stack of the quantum system, bridging device-level and system-level perspectives. Students investigate how abstract quantum circuits are compiled into native operations and further translated into pulse-level control signals. This includes the study of timing, synchronization, and signal generation, providing a system-level understanding of quantum computation workflows.

Experimental activities reinforce both perspectives. Students execute calibration routines, modify pulse parameters, and analyse their impact on measurement outcomes, coherence properties, and gate performance. This enables direct comparison between simulated behaviour and experimental results, highlighting the role of non-idealities and noise in real quantum systems.

From a system design perspective, a major focus is placed on hardware-aware circuit mapping and optimisation. Students are required to map logical circuits onto the constrained topology of the quantum processor, explicitly accounting for connectivity, gate fidelities, and coherence times. These activities expose the trade-offs between circuit depth, SWAP overhead, and execution fidelity, framing compilation as an engineering problem driven by physical constraints.

More advanced activities introduce scalability-oriented techniques such as circuit decomposition and circuit knitting, enabling the execution of circuits exceeding the native hardware capabilities. Students analyse trade-offs between decomposition granularity, reconstruction complexity, and noise accumulation, naturally introducing hybrid quantum–classical workflows and highlighting the limits of near-term quantum hardware.

Finally, selected activities explore the integration of quantum algorithms with classical digital platforms, including FPGA-based emulation and control infrastructures. This enables the study of hybrid architectures, low-latency control, and the co-design of quantum and classical subsystems.

Overall, these activities establish a coherent educational pathway that spans from device-level understanding to system-level optimisation, training students to reason across the full quantum computing stack.

%%%%%%%%%%%%%%%%%%%%%%%%%%%%%%%%%%%%%%%%%%%%%%%%%%%
\section{Related Work}
\label{sec:related}
%%%%%%%%%%%%%%%%%%%%%%%%%%%%%%%%%%%%%%%%%%%%%%%%%%%

\subsection{Vendor cloud platforms} IBM Quantum~\cite{ibmq}, AWS Braket~\cite{braket}, and Azure Quantum~\cite{azure} provide project and quota abstractions similar to what we implemented. However, as proprietary cloud platforms, they offer no insight into internal implementation, cannot be deployed on-premises, and impose vendor-specific access models. Our work implements an analogous management layer for on-premises hardware with full operator control. IQM Resonance~\cite{resonance} is closer to what we propose, in the sense that a subset of its features can be deployed on-premise in the form of IQM Server, which was not available when this work was planned, but it's included in IQM OS 4.4.x; notably, the on-premise version (up to 4.4.6) currently provides only exclusive reservations, while pay-as-you-go budget is not supported.

\subsection{European quantum infrastructure} The EUROQHPC-I Project, funded by EuroHPC, is working on a solution based on similar requirements, that would allow federated access to several EuroHPC-funded quantum computers, respecting site-specific billing and accounting policies. The specific implementation is still being discussed at consortium level and not yet publicly disclosed. Another relevant effort is being made within the Munich Quantum Valley initiative, with the Munich Quantum Software Stack (MQSS), which includes the Quantum Device Management Interface (QDMI)~\cite{qdmi}, which is capable of performing some of the tasks implemented by our platform, and it allows the operator to implement its own site-specific logic with their \emph{driver} component; the limit in this case is the need to involve the vendor in the development of the \emph{device} component, and the need for the user to install a specific client. Also in the context of EuroHPC, Poznan Supercomputing and Networking Centre (PSNC) researchers  developed their own integration software stack that allows both submission from SLURM, and directly via remote API to their own photonic quantum computers~\cite{poznan}.

%%%%%%%%%%%%%%%%%%%%%%%%%%%%%%%%%%%%%%%%%%%%%%%%%%%
\section{Conclusion}
\label{sec:conclusion}
%%%%%%%%%%%%%%%%%%%%%%%%%%%%%%%%%%%%%%%%%%%%%%%%%%%

We have described the design, implementation, and operational experience of the Lagrange management stack, a middleware layer for the IQM Spark quantum computer deployed in a multi-institutional research and education environment in Turin, Italy. The system addresses a gap common to on-premises quantum hardware deployments: vendors provide authentication, but the full resource management, billing, and federation stack must be built by the hosting institution.

Our key design decisions---maintaining full transparency to IQM client tooling, implementing a plugin architecture that separates vendor-specific from site-specific concerns, operating a dedicated Keycloak instance with OIDC/SAML federation, and decoupling result persistence from the synchronous submission path---have proven sound in production. After more than one year of operation and over 850,000 quantum jobs, the system has served a diverse user population ranging from metrology researchers at INRiM to students taking formal examinations at Politecnico di Torino, with no architectural changes required to accommodate these different use cases.

The scale of usage (850,000+ jobs, more than 360 hours of QPU execution time), and the breadth of the user base demonstrate that small-scale quantum computers can be operated as shared, multi-tenant research infrastructure with the same level of service management expected of HPC facilities. In particular, the resource management concepts familiar from HPC centres (project-based allocations, accounting, reservations, and fair-share policies) transfer well to quantum hardware. The fact that students regularly use a real quantum computer during both courses and examinations---a practice we believe to be unique in Europe---illustrates the educational potential that proper management infrastructure unlocks.

The architecture described here provides a reusable blueprint for any institution deploying an on-premises quantum computer in a shared, multi-tenant model. To support its adoption, the whole software stack is now released as open source under the EUPL-1.2 licence: besides QC Gateway, this includes the backend API\footnote{\url{https://github.com/LINKS-Foundation-CPE/Lagrange-API}}, the projects dashboard\footnote{\url{https://github.com/LINKS-Foundation-CPE/Lagrange-Dashboard}}, and the software documentation for developers and operators\footnote{\url{https://links-foundation-cpe.github.io/Lagrange-stack-docs/}}, while the user-facing documentation on accessing Lagrange through its portal is available separately\footnote{\url{http://docs.quantum.linksfoundation.com/}}. A portable deployment kit\footnote{\url{https://github.com/LINKS-Foundation-CPE/Lagrange-stack-portable}} installs the complete stack (QC Gateway, backend API, dashboard, Keycloak, object store, Redis, and PostgreSQL) on a single Linux host with one command, against either a mock device or a real machine, allowing other sites to evaluate the system before adapting the site plugin to their own policies; an optional variant also provisions a containerised SLURM cluster to experiment with HPC-QC integration. We welcome contact from other institutions facing similar deployment challenges.

\ifanon\else
\section*{Acknowledgments}
The authors acknowledge the QTech Piemonte strategic initiative, which was instrumental to create the QCS-Lab and jointly fund the acquisition of the machine by Fondazione LINKS, Politecnico di Torino and Istituto Nazionale di Ricerca Metrologica (INRiM).
The authors also acknowledge the invaluable support of IQM personnel in the process of designing the middleware for the Lagrange machine.
\fi

\ifanon\section*{Generative AI disclosure}\else\subsection*{Generative AI disclosure}\fi
In addition to language and grammar polishing, the authors acknowledge the usage of Claude to draft a skeleton of section~\ref{sec:architecture} from the software documentation, and to provide an initial version of Figure~\ref{fig:sequence}. The text and the figure have been thoroughly reviewed and significantly edited by the authors.

%%%%%%%%%%%%%%%%%%%%%%%%%%%%%%%%%%%%%%%%%%%%%%%%%%%
\bibliographystyle{IEEEtran}
\bibliography{references}

\end{document}